\documentstyle[aps,epsfig,prd,amsfonts]{revtex}

\def\eq#1{(\ref{#1})}

\begin{document}

%@twocolumnfalse\endcsname

\draft

\title{Infrared and ultraviolet cutoffs of quantum field theory}

\author{J.M. Carmona\cite{Email1} and J.L. Cort\'es\cite{Email2}}
\address{Departamento de F\'{\i}sica Te\'orica, Universidad de Zaragoza,
50009 Zaragoza, Spain}

\maketitle
\begin{abstract}
Quantum gravity arguments and
the entropy bound for effective field theories proposed in PRL~{\bf 82},
4971 (1999) lead to consider two correlated scales which parametrize departures
from relativistic quantum field theory at low and high energies.
A simple estimate of their possible phenomenological implications
leads to identify a scale of around 100 TeV as an upper limit on the domain
of validity of a quantum field theory description of Nature. 
This fact agrees with recent theoretical developments in large extra
dimensions. Phenomenological consequences in the beta-decay spectrum and
cosmic ray physics associated to possible Lorentz invariance violations
induced by the infrared scale are discussed. It is also suggested that
this scale might produce new unexpected effects at the quantum 
level.
\end{abstract} 

%%\begin{keyword}
%%Effective field theories; Large extra dimensions; Lorentz invariance; 
%%Beta decay; Cosmic rays
%%\end{keyword}

%% 11.10.-z Field theory
%% 04.50.+h Gravity in more than four dimensions, Kaluza-Klein theory, 
%%unified field theories; alternative theories of gravity
%% 11.30.Cp Lorentz and Poincare invariance
%% 95.85.Ry Neutrino, muon, pion and other elem. particles; cosmic rays  
%% 23.40.-s Beta decay; double beta decay; electron and muon capture
%% 13.85.Tp Cosmic-ray interactions
%]

Local quantum field theory (QFT) 
is a good effective description of Nature at low energies. However, it is
usually assumed that QFTs have a high-energy
limitation of applicability. This is obvious for non-asymptotically free
theories, like QED and scalar $\phi^4$ theory, which seem to fail at very
high energy~\cite{trivialidad}, but it appears to be an inevitable 
general conclusion when trying to incorporate gravity due to the 
non-renormalizability of a QFT of gravitation. 
Moreover, it has been proposed that gravitational stability of the
vacuum sets a limit on the shortest scale of any QFT compatible
with gravity, which is somewhat above the Planck length~\cite{foffa}.

As a consequence, QFTs would be low-energy
approximations to a more fundamental theory that may not be a field
theory at all~\cite{weinberg}, and are valid up to energies
below a certain scale $\Lambda$, which then represents an ultraviolet (UV)
cutoff of the effective field theory. The failure in obtaining a QFT
of gravity indicates that this scale is lower than 
the Planck mass $M_P$, the scale at which the gravitational strength 
is comparable to that of the rest of the fundamental interactions.
Can we tell anything more about the order of magnitude of the scale
$\Lambda$? 

In a QFT, the maximum entropy $S_{\rm max}$ 
scales extensively, with the
space volume~\cite{susskind}, which is a reasonable guess for a local
theory. For a QFT in a box of size $L$ with UV cutoff $\Lambda$,
$S_{\rm max}\sim L^3\Lambda^3$~\cite{cohen}. However, Bekenstein 
arguments~\cite{bekenstein1y2}, 
based on black hole gedanken experiments
and the validity of the generalized second law of 
thermodynamics~\cite{bekenstein3}, lead to think that in a quantum
theory of gravity, the maximum entropy should be proportional to the area
and not to the volume~\cite{susskind}, 
\begin{equation}
S\lesssim L^2 M_P^2 . 
\label{holography}
\end{equation}
Eq.~\eq{holography} is usually called the holography entropy bound.
This bound suggests that conventional field theories
over-count degrees of freedom~\cite{thooft,susskind}, and
implies the breakdown of any effective field
theory with an UV cutoff to describe systems which exceed a certain
critical volume $L^3$ which depends on the UV cutoff. From 
Eq.~\eq{holography}, it is straightforward that
$L \lesssim M_P^2 \Lambda^{-3}$.
This observation lead to
Cohen, Kaplan and Nelson~\cite{cohen} to endow every effective
field theory with an infrared (IR) cutoff correlated to its UV one.
However, using very plausible arguments~\cite{cohen}, they noticed
that conventional QFT should fail at an entropy well below
the holography bound. A QFT should not attempt to describe systems 
containing a black hole. Therefore it should not include states with an
energy greater than $L M_P^2$. 
Since the maximum energy density in
a QFT with UV cutoff $\Lambda$ is $\Lambda^4$, one such a state can be
formed when the size of the system is
\begin{equation}
L \sim \frac{M_P}{\Lambda^2} .
\label{cohensize}
\end{equation}

In Ref.~\cite{cohen}, the scales $L$ and 
$\Lambda$ are not considered as absolute. They signal 
the range of validity of QFT calculations when
applied to a phenomenon of a certain energy scale. However, as noted above,
it is believed that QFT has an absolute UV limit of validity
$\Lambda$ which is probably related to gravitational effects. 
Therefore, this necessarily implies the existence of 
a related absolute IR cutoff $\lambda$ which, according to Eq.~\eq{cohensize},
will be given by
\begin{equation}
\label{relacion}
\lambda \sim \frac{\Lambda^2}{M_P} .
\end{equation}

If one now considers that $\Lambda\sim M_P$, Eq.~\eq{relacion} gives the
absurd result $\lambda\sim M_P$. It is natural to think that $\lambda$ 
should be low enough to make compatible the breakdown of QFT at low
energies with its success in particle physics. This is so however unless
the departures from QFT through effects produced by the scale $\lambda$ were
completely suppressed by a factor proportional to the gravitational
coupling. If this were the case, one
could not derive any relevant bounds for the IR and UV scales. This
scenario is indeed a possibility. However in the following we will assume
that this is not the case, and that the $\lambda$-induced effects
do not have such a suppression factor and may have observable consequences.

Coming back to relation~\eq{relacion}, note that a low
value of $\lambda$ implies a limit of validity of QFT at high energies 
much lower than the Planck mass $M_P$. 
Precision tests of the electroweak standard model put the lower bound for
$\Lambda$ in the TeV range. On the other hand, as we will see below, a
value of $\lambda$ higher than the scale of eV (and even smaller values 
depending on the assumptions used to estimate the effects of the IR
scale) could have observable effects in certain experiments, like the
tritium beta-decay. These two observations restrict very much the 
possible ranges of the two cutoffs: $\Lambda\sim (1 - 100)$ TeV and
$\lambda\sim (10^{-4} - 1)$ eV. 
Therefore QFT would be an effective theory valid only up to an
energy scale of the order of 100 TeV.
If one takes this cutoff as the fundamental short distance scale then one
has the possibility to explore a new framework for solving the hierarchy
problem which does not rely on either supersymmetry or technicolor.
A disadvantage of this scenario is that one would loose the successful
prediction of grand unification theories for $\sin \theta_{\rm W}$,
but anyway the correct prediction might come out in the end in a more
complicated way.

If, as it is commonly accepted, the origin of the 
limitation at high energies of QFT is the gravitational
interaction, it is natural to identify $\Lambda$ with the fundamental
scale of gravity. In fact, the existence of a fundamental scale of the
gravitational interaction well below the Planck mass scale, and just
a few orders of magnitude above the electroweak scale, a very attractive
idea to solve the hierarchy problem, has been proposed in some recent
innovative works~\cite{arkanietal,cohen2} on
extra dimensions. In the approach of large extra
dimensions~\cite{arkanietal}, the observed
hierarchy between the electroweak and Planck scales is explained by
postulating a fundamental scale $M\sim 10 - 100$ TeV of gravity 
along with Kaluza-Klein compactification with large radius $R$. The Planck
scale is then an effective four-dimensional scale.
The case of two extra
dimensions is particularly interesting. In this case, the radius of the
compact extra dimensions is given by
\begin{equation}
\label{extradim}
R \simeq \frac{M_P}{M^2} , 
\end{equation}
which is close to the present limit of validity of Newton's
Gravitational Law. It is surprising to note the similarities between  
Eqs.~\eq{relacion} and~\eq{extradim} if one identifies the UV limit
with the fundamental scale of gravity in $4+2$ dimensions ($\Lambda\sim M$) 
and the IR limit with the inverse of the radius of compactification of 
the large extra dimensions ($\lambda\sim 1/R$), a relation to which we find
no simple interpretation.
Alternatives to the hierarchy problem in terms of finite, but non-compact
extra dimensions require also a fundamental scale of the order of
100 TeV~\cite{cohen2}.

In the following we will explore both the phenomenological and
theoretical consequences of these stringent limits of a QFT
description of Nature, and the compatibility of these limits 
(and therefore, of the arguments which lead to Eq.~\eq{cohensize}) with 
present experimental status.

Let us examine first the deviations from QFT at low energies.
We will consider two different scenarios for the effects of the IR 
scale. In the first scenario we will assume that the neutrino is particularly 
sensitive to the presence of the IR scale. This is a natural 
assumption in the framework of large extra dimensions where the neutrino 
is the only particle (besides the graviton) which propagates in the 
extra dimensions. Then any dependence on the IR scale (compactification 
radius) for the remaining particles requires to consider the gravitational 
coupling and will then be suppressed.
For an IR cutoff of the order of the eV, effects should be seen in
experiments sensitive to neutrinos with energies in the $10-100$ eV scale. 
Indeed, it is in this range where recent experiments of the tritium 
beta-decay have observed an
anomaly which consists in an excess of electron events at the end of the
spectrum, at about 20 eV below the end point~\cite{lobashev}.
A value of $\lambda$ higher than few eV would produce
a signal in the tritium beta-decay spectrum in a larger range than that
of the observed anomaly, so this gives the bound $\lambda < 1$ eV in 
this scenario, which produces the severe limit $\Lambda < 100$ TeV on the
UV cutoff of QFT. But besides putting an upper limit on $\lambda$, the 
tritium beta-decay spectrum could also allow to identify a correction 
induced by the IR cutoff. Indeed a modification of the dispersion
relation for the neutrino, of the form
\begin{equation}
E^2 = \bbox{p}^2 + m^2 + \mu |\bbox{p}| ,
\label{lorentz1}
\end{equation}
has been used to explain the tritium beta-decay anomaly~\cite{tritio}. 
Matching with experimental results requires a value of $\mu$ in 
Eq.~\eq{lorentz1} of the order of the eV, 
and it can be seen that this does not 
contradict any other experimental result~\cite{tritio}. 

In fact a deviation from QFT at low energies due to the IR 
scale $\lambda$ may well be expected
to violate relativistic invariance. The reason is that it has been 
shown~\cite{weinberg} that {\it any} theory incorporating quantum mechanics
(QM) and special relativity, with an additional ``cluster'' 
condition~\cite{weinberg}, must reduce to a QFT at low energies.
A modified dispersion relation of the form of Eq.~\eq{lorentz1} 
is a simple way of incorporating effects beyond QFT 
that violate relativistic invariance at low energies. 
Note that, together with the dispersion relation~\eq{lorentz1},
one should indicate the ``preferred'' frame in which this relation
is valid. There is however another possibility, which is to extend our
concept of relativistic invariance to a more general framework, in
which Eq.~\eq{lorentz1} would be an observer-invariant relation.
The possibility to have a modification of Lorentz transformations
compatible with the presence of an observer-independent scale
of length has been only recently initiated to be 
explored~\cite{amelino}.

In order for the dispersion relation Eq.~\eq{lorentz1} to be compatible with
the very stringent limits on CPT violation~\cite{CPT} it is necessary to 
have the same scale $\mu$ in the particle-antiparticle dispersion relation.
Even with this limitation a modified dispersion relation for any particle 
with $\mu \sim \lambda$ is not compatible with experimental limits.
For example, Eq.~\eq{lorentz1} for the electron 
would slightly modify the energy levels of the hydrogen atom. 
Given the extraordinary agreement between theory
and the experimental measurement of the Lamb shift (one part in 
$10^5$~\cite{kinoshita,weinberg}), one has $\mu<10^{-6}-10^{-7}$ eV.
Therefore, an identification of the scale $\mu$ which parametrizes the
Lorentz invariance violation (LIV) at low energies in the dispersion
relation with the IR scale $\lambda$ is incompatible with the arguments
which lead to Eq.~\eq{cohensize}. A way to reconcile these 
arguments with a LIV at low energies is in the framework of the scenario
described above, in which one expects a suppression of the dependence of 
LIV effects on the IR scale for all particles except for the neutrino. 
This would explain why no signal of these LIVs has been observed. 
In the case of the neutrino no such suppression is present
and then $\mu \sim \lambda$; besides that the neutrino mass is not larger than 
the IR scale and this makes possible to observe the consequences of a
Lorentz noninvariant term in the neutrino dispersion relation.
Then the anomaly in tritium beta-decay, if confirmed as a real physical 
effect, could be the first manifestation of the IR cutoff of QFT. 
If the anomaly results to be a consequence of some systematic 
effect not taken into account~\cite{lobashev} then one can obtain a
stronger upper limit on the IR cutoff. 

We also note that the new dispersion relation 
Eq.~\eq{lorentz1} might have important effects
in cosmic rays through threshold effects which become relevant when
$\mu |{\bbox p}_{\rm th}|\sim m^2$. Here $m^2$ is an ``effective''
mass squared which controls the kinematic condition of allowance or 
prohibition of an specific process. Indeed a consequence of 
these threshold effects could be that neutrons and pions become 
stable particles at energies close to the knee of the cosmic ray 
spectrum~\cite{tritio}, which would drastically alter the composition 
of cosmic rays. It is quite remarkable that cosmic ray phenomenology
could be sensitive to the presence of an IR scale.

Since the IR scale was introduced by general arguments reflecting 
the apparent incompatibility of QM with a complete theory which
contains gravity, it is natural to consider a
second scenario in which the effects of the IR scale are due to the 
quantum fluctuations of the vacuum and then affect all the particles. 
In this scenario the most stringent limits on 
the IR scale come from the high precision tests of QED, in particular
from the anomalous magnetic moment of the electron. In this case, the 
characteristic physical scale is the electron mass, but the precision 
achieved makes the experiment sensitive to much lower scales. In fact, 
it gives the most precise test of QED. We should therefore ask whether 
it is compatible with the presence of an IR cutoff.
A simple estimate of the correction to the usual calculation imposed
by the IR scale is
\begin{equation}
\label{correccion}
\delta a_e \sim \frac{\alpha}{\pi} \left(\frac{\lambda}{m_e}\right)
\sim 4\times 10^{-9} \frac{\lambda}{(1 \ {\rm eV})}.
\end{equation}
If we ask this correction to be smaller than the uncertainty of the 
theoretical prediction for $a_e$ in QED caused by the uncertainty in the 
determination of $\alpha$~\cite{kinoshita}, we get a bound for the IR 
scale $\lambda \lesssim 10^{-2}$ eV. Other tests of QED, like shifts of 
energy levels in hydrogen atom, positronium, etc, lead in this case 
to less stringent bounds on the scale $\lambda$. 
The previous bound on the IR scale
corresponds to an UV scale $\Lambda \lesssim 10$ TeV which is very close 
to the present and near future energies available in accelerator physics.

This second scenario suggests that QM, as we know it
today, might fail not only above the 10 TeV scale, 
but also that one could find 
unexpected effects at the quantum level for phenomena with a 
characteristic scale of $10^{-2} - 10^{-4}$ eV, for example diffraction 
experiments with wave-lengths of the order of a millimeter. 
Quantum systems which are sensitive to wave-lengths of this order of magnitude
are candidates to show departures of QFT parametrized by the IR scale. 
The understanding of the transition between the quantum and 
classical regimes, going from a QFT description with
departures parametrized by an IR scale to the classical theory, would require 
the use of the more fundamental theory which could provide a solution to the 
quantum mesurement paradox~\cite{bell}.
Finally, we mention that some aspects of the large scale structure of 
the Universe which are related to quantum fluctuations in an early epoch 
of its evolution could also include signals of the IR cutoff. 

We turn now our attention to possible departures from QFT results
for high-energy phenomenology above the $1-100$ TeV limit. 
A natural candidate to reveal new physics beyond this energy scale is
cosmic ray physics, where several anomalies are observed~\cite{olinto,watson}.
However, a detailed discussion of the expected effects requires specific 
formulations of the kind of limitations presented by QFT, like the ones 
explored in the context of extra dimensions~\cite{extradimphenom}.
 
Generally speaking, the presence of a cutoff $\Lambda$ induces also
non-renormalizable corrections to the effective field theory. For example,
conventional QED would include modifications produced by
a Lorentz, gauge, and CP invariant Pauli term of dimension 5, 
of order $1/\Lambda$ or, considering extra
symmetries that restrict the form of the non-renormalizable interactions,
of order $m/\Lambda^2$~\cite{weinberg}. In the first case, the theoretical
and experimental agreement on the value of the magnetic moment of electron
gives $\Lambda \gtrsim 4\times 10^7$ GeV, which would mean the invalidity
of the arguments that lead to Eq.~\eq{cohensize}. In the second case,
it is the value of the muon magnetic moment~\cite{brown} rather than that
of the electron the one which provides the most useful
limit on $\Lambda$, $\Lambda\gtrsim 3\times 10^3$ 
GeV~\cite{weinberg}.\footnote{The Muon ($g-2$) Collaboration has recently
reported (hep-ex/0102017) a possible incompatibility between the
experimental value of the muon magnetic moment and its theoretical value
from the standard model. This difference could be explained by
non-renormalizable corrections induced by the presence of an ultraviolet
cutoff $\Lambda$ around $4-5$ TeV.}
This is why it is usually considered that conventional QFT gives a 
correct description of Nature at least up to 
the scale of the TeV.\footnote{The less stringent limit on 
$\Lambda$ can also be
understood without the need of additional symmetries in the framework
of the two considered scenarios, in which the Pauli term will be
suppressed, either by the gravitational scale in the first scenario,
or by quantum fluctuations in the second.}
This lower bound for $\Lambda$ is the origin of the bound
$\lambda\gtrsim 10^{-4}$ eV on the IR cutoff.

Let us also consider the possibility of LIVs at high energies. 
Several attempts have been made to question Lorentz 
invariance~\cite{kostelecky}.
The existence of LIVs at high energies is natural
in the context of quantum gravity~\cite{jacobsonythooft2,amelinocamelia1}. 
Quantum gravity
fluctuations produce in general modifications in the dispersion relations
which characterize the laws of particle propagations
\cite{amelinocamelia1,gambiniyamelinocamelia2ymorales,amelinocamelia3,coleman}.
These modifications have been used~\cite{amelinocamelia3,coleman} to 
explain the observed violations of the GZK cutoff limit~\cite{GZK}.
In fact, it has been recently shown~\cite{GZKyMk} that these violations of
Lorentz invariance can at the same time offer a solution to the anomaly
observed~\cite{aharonian} in the gamma-ray spectrum of Markarian 501,
which extends well beyond 10 TeV.

As an example, let us consider the Lorentz-violating class of
dispersion relations~\cite{amelinocamelia3}
\begin{equation}
E^2 = {\bbox p}^2 + m^2 + \frac{|{\bbox p}|^{2+n}}{M^n},
\label{lorentz2}
\end{equation}
where $M$ is the characteristic scale of these violations
(the fundamental scale of gravitation in the quantum gravity framework).
It is then easy to see that $M$ causes the appearance of threshold effects at
momenta $|{\bbox p}|\gtrsim |{\bbox p}_{\rm th}|$, where 
$|{\bbox p}_{\rm th}|^{2+n}\sim m^2 M^n$. It is these effects which allows
to explore quantum gravity at energies much lower than $M$~\cite{grillo}.
For $M\sim M_P$ and a typical hadronic process, 
one gets $|{\bbox p}_{\rm th}|\sim 10^{15}$ eV in the case $n=1$ and
$|{\bbox p}_{\rm th}|\sim 10^{18}$ eV in the case $n=2$. In both cases one
has modifications to relativistic kinematics at energies below the 
GZK cutoff, so that the observed violations of this 
cutoff in the cosmic ray spectrum~\cite{watson} could be a footprint of
a LIV at high energies. The fact that these violations can also 
offer a solution to the Markarian 501 anomaly is easily seen 
if one considers the threshold of the reaction
$\gamma +\gamma \to e^+ e^-$, which restrains
the propagation of gamma rays in the IR background. The effective
mass that controls this process is $m^2\sim 1$ MeV$^2$, so that
$n=1$ gives $|{\bbox p}_{\rm th}|\sim 10$ TeV, and the conventional threshold
is modified.

The characteristic scale $M$ for LIVs at high energies does not have
to coincide in principle with the UV scale $\Lambda$, which is
defined as the maximum energy of the quantum fluctuations whose effects
can be described using QFT, though $M$ will surely depend on $\Lambda$.
In fact, a scale $M$ much larger than $\Lambda$ can be
justified for all particles except for the 
neutrino in the first scenario if one assumes 
that the suppression of the dependence on the IR scale applies also  
to the dependence on the UV scale. On the other hand in the case 
of the neutrino one expects that $M \sim \Lambda$ and then more clear
signals of LIVs at high energies in reactions involving neutrinos.
Alternatively, in the second scenario all the effects of the IR and 
UV scales are due to the quantum fluctuations of the vacuum and 
then one expects $\mu \ll \lambda$ and $M \gg \Lambda$ for the scales that
parametrize LIVs at low and high energies for any particle.  
In both scenarios the scale $M$ can be made sufficiently large to be 
consistent with the tight constraints from experiments on Lorentz and CPT
violations~\cite{kostelecky2,coleman}.  

In conclusion, quantum gravity, Bekenstein's and
Cohen, Kaplan and Nelson's arguments, together with their phenomenological
consequences, indicate that a QFT description of Nature
might not be valid above the scale $\Lambda\sim (1 - 100)$ TeV and that  
departures could be seen at low energies characterized by an IR
scale $\lambda\sim (10^{-4} - 1)$ eV.
Interpreting $\Lambda$ as the fundamental scale of gravity,
the surprise is that the UV limit of QFT would not be the Planck scale, 
but 14 orders of magnitude lower. This would made
quantum gravity phenomenology much more accessible, and agrees with 
recent theoretical development on extra dimensions. We have identified 
certain experiments that could reflect the limitations of QFT and explored
in which scenarios these ideas are compatible with the present
experimental status quo.

As a final comment, we would like to speculate with the hypothesis that
the apparition of two correlated scales in QFT might be a general property
of every extension of QFT which tried to incorporate the gravitational
interaction. An example is the IR/UV connection in non-commutative gauge 
theories~\cite{matusis} which arise as effective field theories of the 
tachyon and gauge field degrees of freedom of the open string in the 
presence of D-branes~\cite{seiberg}. 
Another example is large extra dimension models, where,
besides the fundamental scale of gravity, one has to introduce another
scale corresponding to the compactification radius of the extra dimensions.
The presence of two correlated scales is particularly suggestive to 
give an answer to the cosmological
constant problem, which seems to require a correlation between the Fermi
energy scale, 300 GeV, and the cosmological constant scale, $10^{-2}$ eV,
in order to explain the enormous precision in the cancellation of the
vacuum energy density contribution of the standard model. It is 
noticeable that these two scales coincide very approximately with the
UV and IR scales of QFT that we have identified through
phenomenological arguments.

We are grateful to Stefano Foffa and J.G. Esteve 
for useful discussions.
The work of JMC was supported by EU TMR program
ERBFMRX-CT97-0122 and the work of JLC by MCYT (Spain), grant
FPA2000-1252.

\end{document}